# Robust singlet dimers with fragile ordering in two-dimensional honeycomb lattice of Li$_2$RuO$_3$

Junghwan Park*[1], Teck-Yee Tan*[2], D. T. Adroja[3,4,$], A. Daoud-Aladine[3], Seongil Choi[1,2], Deok-Yong Cho[1,2], Sang-Hyun Lee[1,2], Jiyeon Kim[1], Hasung Sim[2,5], T. Morioka[6], H. Nojiri[6], V. V. Krishnamurthy[7], P. Manuel[3], M. R. Lees[8], S.V. Streltsov[9,10], D.I. Khomskii[11,$], and Je-Geun Park[1,2,5,$]

[1] Center for Strongly Correlated Materials Research, Seoul National University, Seoul 08826, Korea
[2] Center for Correlated Electron Systems, Institute for Basic Science, Seoul 08826, Korea
[3] ISIS Facility, Rutherford Appleton Laboratory, Didcot OX11 0QX, United Kingdom
[4] Highly Correlated Matter Research Group, Physics Department, University of Johannesburg, Auckland Park 2006, South Africa
[5] Department of Physics and Astronomy, Seoul National University, Seoul 08826, Korea
[6] Institute for Materials Research, Tohoku University, Sendai 980-8577, Japan
[7] Department of Physics and Astronomy, George Mason University, Fairfax, VA 22030-4444, USA
[8] Department of Physics, University of Warwick, Coventry CV4 7AL, United Kingdom
[9] Institute of Metal Physics, Ekaterinburg 620041, Russia
[10] Department of Theoretical Physics and Applied Mathematics, Ural Federal University, Ekaterinburgh 620002, Russia
[11] II Physikalisches Institut, University of Koeln, 50937 Koeln, Germany

* Equally contributed
$ Correspondence to J.G.P. [email: jgpark10@snu.ac.kr], D.I.K. [email: khomskii@ph2.uni-koeln.de] & D.T.A. [email: devashibhai.adroja@stfc.ac.uk]

When an electronic system has strong correlations and a large spin-orbit interaction, it often exhibits a plethora of mutually competing quantum phases. How a particular quantum ground state is selected out of several possibilities is a very interesting question. However, equally fascinating is how such a quantum entangled state breaks up due to perturbation. This important question has relevance in very diverse fields of science from strongly correlated electron physics to quantum information. Here we report that a quantum entangled dimerized state or valence bond crystal (VBC) phase of Li$_2$RuO$_3$ shows nontrivial doping dependence as we perturb the Ru honeycomb lattice by replacing Ru with Li. Through extensive experimental studies, we demonstrate that the VBC phase melts into a valence bond liquid phase of the RVB (resonating valence bond) type.

This system offers an interesting playground where one can test and refine our current understanding of the quantum competing phases in a single compound.

Systems with strongly correlated electrons usually harbour rich magnetic properties[1]. Most often these are different types of long-range magnetic ordering. However, other options are also possible. One of them, widely discussed and still being extensively studied, is the formation of different types of spin liquid states, which is generally expected to be realised in frustrated systems[2]. Yet another possibility is that singlet bonds are formed in the system, leading eventually to the emergence of valence bond crystals or valence bond solids. These are, for example, Peierls and spin-Peierls states in one-dimensional systems, but such states can also exist in higher dimensions.

The exact conditions in which such a VBC can be formed are not well known, although some general trends have been noted. One likely possibility is their formation in low-dimensional systems[3]. Orbital degrees of freedom may also help to stabilize such states[4,5] via a particularly intriguing mechanism of orbital-selective dimerization[6]. Although interesting in its own right, the details of how the VBC are formed and stabilized remain largely unexplored. Equally fascinating questions are how the crossover between the usual magnetic ordering and the VBC occurs, and what the possible metastable states are close to such a crossover. There are in principle two possibilities: either there exist a quantum phase transition between these states, or the transition between them could be of first order. In the latter case, one could expect a possible coexistence of both states, which can imply, in particular, the existence of local dimers close to such transitions.

Another equally interesting and yet poorly understood question is how the VBC with a charge gap responds to low levels of doping. We are not aware of any studies of such a crossover or of the detailed doping effects in a real material under controlled conditions. We now identify one system which seems to be ideal for this kind of study: $Li_2RuO_3$, a layered material with a honeycomb lattice.

Systems with a honeycomb lattice have recently attracted special attention. The most celebrated case is of course graphene, with its Dirac points in the electronic spectrum. Correlated systems with honeycomb lattices are also generating considerable interest. The best known examples are $Li_2IrO_3$ and $Na_2IrO_3$, for which the applicability of a Kitaev-Heisenberg model was proposed[7,8], with the eventual formation of a special spin-liquid state. Unfortunately, however, different magnetic ordered states were found in real materials: a zigzag-type ordering in $Na_2IrO_3$[9] and an incommensurate magnetic ordering in γ-$Li_2IrO_3$[10]. We note that other types of magnetic ordering were also found in related compounds with similar structures: $Li_2MnO_3$[11,12] and $Li_2RhO_3$[13]. However, at least in one system of this class, $Li_2RuO_3$, the situation is drastically different. It was shown by Miura *et al.*[14,15] that in polycrystalline $Li_2RuO_3$ there is a transition of the singlet Ru-Ru dimer formation at $T_c \sim 540$ K, below which these dimers order in a herringbone fashion. Thus this material could be a classic example of the VBC state. As argued in Ref. 16, orbital ordering seems to play a crucial role in the formation of this VBC state.

The situation in $Li_2RuO_3$ is far from trivial and much richer than originally thought. On the one hand, the singlet dimers with short Ru-Ru bonds seem to be very stable. As demonstrated by recent total scattering experiments and PDF (Pair Distribution Function) analysis[17], the Ru-Ru dimers survive up to temperatures much higher than its $T_c$. On the other hand, a recent study of $Li_2RuO_3$ single crystals reported quite different behaviour. According to this latter study[18], depending on the exact preparation conditions some single crystals show only a much weaker transition at about the same temperatures as reported in Ref. 17 whereas other crystals show no transition at all, apart from a weak magnetic ordering at ~ 5 K[18]. Exact reasons for the sample dependent behaviour are not known at the moment, but one can guess that it may be due to small deviations in stoichiometry, which is more difficult to control in single crystals grown using the flux method than in powders.

To have a better understanding of the situation, one has to answer the questions of how the VBC ground state evolves upon doping or introducing disorder in a real 2D material and how the magnetic ground state eventually emerges. This is a nontrivial problem and warrants careful study. In this work, we carried out a detailed study of this system with controlled changes of disorder and employing a comprehensive set of experimental techniques: resistivity, magnetization, specific heat, and both high-resolution elastic and inelastic neutron scattering (see Methods). On the basis of these studies, we conclude that even small levels of disorder, which may be expected to have a negligible effect on the ground state, are found to drastically modify the long-range ordered spin dimer state and thereby influence many of the physical properties of this system. Of particular interest is that all this happens despite the fact that the singlet dimers seem to survive locally even at higher doping. Thus the resulting state may be visualized as some kind of a dimer liquid, but with certain dimers broken. This, in particular, produces unusual spatial modulations in the hopping integrals leading to a variable-range hopping (VRH)-like conduction at low temperatures. Another nontrivial effect observed experimentally is the appearance of low-energy excitations giving rise to linear specific heat at low temperatures: $\frac{C}{T} = \gamma + \beta T^2$, with a large coefficient $\gamma$. The broken dimers also seem to lead to enhanced magnetic signals, observed in the uniform susceptibility measured by bulk magnetization and the dynamic susceptibility obtained from inelastic neutron scattering experiments. Our *ab-initio* GGA (generalized gradient approximation) calculations support this picture and, in particular, show the formation of a magnetic cloud close to impurities (in our case extra Li on the Ru sites).

At high temperatures $Li_2RuO_3$ with $Ru^{4+}$ forms in the *C2/m* space group, one of several monoclinic phases common to this type of transition metal oxides, with the Ru occupying the symmetric honeycomb lattice with Li at the centre of Ru hexagons (Figs. 1a&b). These Ru honeycomb layers are separated by another layer of Li atoms (Fig. 1b). Upon cooling below 550 K, $Li_2RuO_3$ undergoes a structural transition into another monoclinic phase of *P2$_1$/m* symmetry by losing the face centring of the monoclinic plane. As shown in Figs. 1c&d, this transition is accompanied by a strong off-centring of Li atom breaking the symmetric honeycomb networks, and by splitting the three otherwise equal Ru–Ru nearest neighbour distances on the honeycomb lattice into one short and one intermediate and one long bond per unit cell (Supplementary Fig. 1 and Table 1). This off-centring and lowering of the symmetry was originally interpreted as due to Ru dimerization coupled with magnetoelastic coupling with the orbital configuration shown in Fig. 1e[16]. Also notable is the change of the unit cell parameters, and in particular the shrinkage of the lattice in the direction of the short Ru-Ru dimers, thus making the *a (b)* lattice constant shorter (longer), respectively (Supplementary Fig. 1 and Table 1)

All the physical properties of our samples prepared as described in the Methods section exhibit drastic changes at the structural transition as shown in Fig. 2. For example, the resistivity shows a marked increase, reminiscent of a metal-insulator transition seen in other Ru oxides[19]. At the same time, the magnetic susceptibility displays a sudden drop at almost the same temperature as reported previously[14,15]. What is intriguing is that the low temperature resistivity of all our samples does not follow the usual Arrhenius law although the high temperature phase seems to be more consistent with the activation behaviour with a charge gap energy of around 200 meV (see Fig. 2a). The low temperature behaviour is clearly closer to variable-range hopping behaviour. Another interesting feature is that the magnetic susceptibility shows persistent paramagnetic signals of $2.5\times10^{-4}$ emu/mole over a very wide temperature range of almost 500 K below the transition, implying a nonzero density of states in the spin susceptibility (see Fig. 2b). This then appears to be at variance with the picture of a complete dimerization as suggested earlier[14,15].

The Ru honeycomb lattice can be perturbed by introducing more Li atoms into the Ru honeycomb lattice: for example, one can replace Ru by Li and vice versa. An extreme example of such disruption of the Ru honeycomb lattice is $Li_3RuO_4$, which has one-dimensional zig-zag chains of Ru with $Ru^{5+}$ valence, and is known to order antiferromagnetically at 40 K[20]. In order to investigate disorder effects on the metal-insulator transition (MIT) of $Li_2RuO_3$, we have made careful experiments, controlling the levels of disorder of the Ru honeycomb lattice by replacing Ru with Li ions and vice versa, i.e. mixing between Ru and Li atoms. In order to determine the exact amount of mixing ($x$) we carried out high-resolution neutron/X-ray diffraction experiments (Supplementary Table 4) and compared it with our best sample (LRO1 with a nominal value of $x \cong 0$) to find that there is a monotonous variation in the unit cell volume with the amount of disorder for the $x$ values of our interest. This picture of disorder at the Ru honeycomb lattice by Li is confirmed by the observation that as we increase the mixing ratio ($x$) between Ru and Li in the Ru honeycomb lattice with more interchange between Li and Ru, the lattice parameter $a$ increases while the lattice parameter $b$ decreases (Supplementary Table 3).

As shown in the resistivity and magnetization data in Figs. 2a&b, the transition is progressively suppressed with increasing $x$. Several things are noteworthy here. First, the transition temperature does not change much while the transition itself becomes significantly broadened with doping. Then, the variable-range hopping behaviour seen below the transition temperature of $Li_2RuO_3$: $\rho = \rho_o \exp(\frac{\Delta}{kT})^\alpha$ with the exponent α ~ 1/3 and the charge excitation gap of Δ = 30 K for $x = 0$, transforms into a more insulating state with an activation energy increasing with $x$ (Fig. 2a). For comparison, there is a clear activation behavior with Δ = 320 K for $Li_3RuO_4$ (Fig. 2a). We note in passing that the magnetoresistance of $Li_2RuO_3$ measured up to 14 tesla is always positive.

Simultaneously, the drop in the susceptibility becomes less evident with increasing disorder, before recovering behaviour more reminiscent of Curie-Weiss localized moments for the $Li_3RuO_4$ sample, which shows the antiferromagnetic transition at 40 K as reported previously. We comment that according to our high-resolution X-ray diffraction data there is a sign of a miscibility gap for samples with higher disorder than reported here.

As regards the disorder effects on the MIT, it is worth mentioning that all our samples, including a sample ($x$ =0.13) labelled DTA that was prepared under slightly different conditions (see Methods), show the metal-insulator transition with more or less the same activation behaviour above the transition while there are clear variations in their low temperature behaviour (Fig. 2a). This sample dependence of the low temperature resistivity indicates that the low temperature phase is rather sensitive to low levels of disorder, while the transition itself appears to be more robust to modest doping. This experimental observation is rather remarkable given the high transition temperature.

What is more striking is the low temperature behaviour of the heat capacity shown in Fig. 2d. For our most stoichiometric sample (LRO1 with $x = 0$) with the highest transition temperature, the low temperature heat capacity includes a very small but finite linear contribution, with a γ value of 0.87 mJ/mole-$K^2$. With increasing the mixing ratio ($x$), the γ value increases monotonically and reaches a large value of 40 mJ/mole-$K^2$ for $x = 0.22$. For comparison, we measured $Li_2TiO_3$ prepared under similar conditions and found that it has the γ value of 0.09 mJ/mole-$K^2$. Therefore, this measured γ value of 40 mJ/mole-$K^2$ for $x = 0.22$ suggests that the low temperature phase should be a highly correlated phase. To put this value into perspective, let us assume that this γ value directly scales with the degree of disorder (which is confirmed by our data, Fig. 3) and that for $x$=0.22 it comes from ~20% of the total volume of the sample. To put it in perspective and to stress that this γ value is indeed very large, we can consider a hypothetical case of 100% doping: of course, not realizable in practice. It helps one have a better feeling of the largeness of the observed γ values. For such a hypothetical 100%-doped case, the rescaled γ value would be 200 mJ/mole-$K^2$, which is on par with the typical values for heavy fermion systems[21].

In Fig. 3, we summarize the main experimental findings (transition temperature, the gap value estimated from the high temperature resistivity data, the γ value, and the paramagnetic susceptibility contribution) as a function of the unit cell volume: the unit cell volume is found to increase almost linearly with the mixing ratio ($x$), thus these plots can be taken to show the dependence of these parameters with $x$. This summary reinforces our view that the transition temperature is only slightly suppressed with doping while the γ value and the room-temperature susceptibility increase noticeably.

The detailed origin of the most striking observation: linear specific heat, deserves special discussion. As noted above, the specific heat at low temperatures in modestly doped samples (up to $x \sim 0.2$) shows a linear temperature dependence: $C = \gamma T$, with a rather large value of γ. This kind of linear temperature dependence in specific heat is usually ascribed to the presence of the Fermi-surface, with low-energy (electronic) excitations[22] with a finite density of states: $\gamma \sim N(E_F)$, where $E_F$ is the Fermi energy. The observed linear specific heat data at low temperatures could be due to the presence of electronic excitations, e.g. the formation of inhomogeneous (phase-separated) state with *metallic* droplets immersed in the insulating matrix. We cannot exclude this possibility entirely at the moment although it seems rather unlikely considering all the experimental observations, in particular the huge value of γ anticipated for such an imaginary metallic state occupying the whole sample as discussed above.

However, this is not the only possibility. Some other excitations with this property (a finite density of states at zero energy) could also give rise to such linear contributions to specific heat. This is indeed, for example, the case in some disordered systems[23]. In this sense, we can offer an alternative explanation, which is to attribute this linear specific heat to some other excitations in the magnetic subsystem. The ground state of Li$_2$RuO$_3$ consists of the ordered arrangement of singlet dimers, i.e. it is a valence bond crystal (VBC). The total scattering experiments with the PDF analysis[17] demonstrate that at $T>T_c$, i.e. above the structural transition, dimers still exist locally, forming something like a (classical) dimer liquid. One can assume that similar state can also be generated at low temperatures by certain local disorder, i.e. by extra Li replacing some of the Ru ions in the honeycomb layers. In such a state, there should exist real magnetic (singlet-triplet) excitations, contributing to the magnetic susceptibility, but these may have a rather large energy gap (the singlet binding energy). But it is plausible that random and dynamic distribution of the dimers also allow for singlet excitations. It is known, for example, that in some frustrated magnetic systems, e.g. in Kagome magnets, there exist a lot of low-energy singlet excitations, which are *accumulated* at zero energy with increasing system size[24]. We can assume that similar excitations may also be possible in our disordered $Li_2RuO_3$ samples with suppressed long-range dimer ordering, but with dimers surviving locally, and that such excitations can make a significant contribution to the linear specific heat (see also the discussion below, after presentation of neutron scattering data in Fig. 4). But, the details of this picture are still unclear, and we cannot exclude that at least part of this specific heat comes from real electronic excitations.

Further insight into the nature of the disorder-induced state can be obtained from measuring the excitations directly by using techniques such as inelastic neutron scattering on $Li_2RuO_3$ samples with different amount of disorder. For that purpose, we have measured the spin dynamics of two samples with different Li contents, i.e. a different amount of disorder: one is a sample with less disorder (LRO2 with $x$ ~0.07), and the second is a slightly more disordered sample with $x$ ~ 0.13 (DTA) (see Methods). As shown in Fig. 4a, the inelastic neutron scattering data of the DTA sample measured at 5 K exhibits strong scattering over the energy range from 2 to 6 meV. On the other hand, this scattering is strongly suppressed in the data taken on the LRO2 sample ($x$ ~ 0.07) with less disorder as shown in Fig. 4b. That is, the low energy excitations observed in our inelastic neutron scattering experiments are clearly induced by a small amount of disorder, i.e. Ru on the Li site. This conclusion is further supported by the difference taken between the two data sets (Fig. 4c), which is obtained by directly subtracting the LRO2 data from the DTA data.

We note that our subsequent measurement on another sample (LRO5 with $x$~0.16) reproduces exactly the same behaviour of strong magnetic scattering, as shown in the supporting information (Fig. SI5): the LRO5 sample was prepared under a more control protocol as described in the Methods section. Integrating the inelastic neutron scattering data of the DTA sample over the first Brillouin zone suggests that approximately $0.5\mu_B$/f.u. of magnetic moments are involved in the low energy excitations. It ought to be noted that as shown in Fig. SI6, these magnetic excitations are significantly weakened with increasing temperature, although they are still visible even in the data taken at room temperature.

Of further interest is that the low-energy excitations demonstrate a clear momentum modulation, which can be explained by the nearest neighbour correlation as shown in the supporting information (Fig. SI7). These correlations are probably connected with dimer correlations as seen in the total scattering measurements[17]. The other point worth mentioning is that the uniform susceptibility calculated from the inelastic neutron scattering data is in good agreement with the bulk data as shown in the supporting information (Fig. SI8). This latter observation reinforces our view that the unusually enhanced low-temperature susceptibility and the γ value are intrinsic and arise from the low-energy excitations measured by our inelastic neutron scattering experiments. A further confirmation can be found in the so-called Wilson plot as shown in Fig. SI9, the Sommerfeld-Wilson ratio $R = \frac{4\pi^2 k_B^2 \chi_0}{3(g\mu_B)^2 \gamma}$ is less than one for our $Li_2RuO_3$ materials, implying that strong correlations are present in our samples.

The picture emerging out of the magnetic behaviour of $Li_2RuO_3$ with disorder resembles that of the lightly doped canonical spin-Peierls system $CuGeO_3$. Doping of $CuGeO_3$ by nonmagnetic Zn breaks some of singlet pairs, and the resulting unpaired spins polarize the remaining singlet dimers, eventually leading to an inhomogeneous magnetic order[25,26]. We can expect similar behaviour to occur in $Li_2RuO_3$, with the key difference being that because of a very large binding energy of singlet dimers (in our case $T_c$ ~ 500 - 600 K, instead of $T_c$~14 K in $CuGeO_3$), the extension of the *magnetic cloud* around impurities would be much smaller in $Li_2RuO_3$.

To check this hypothesis we performed *ab-initio* GGA calculations and simulated the 4% Li/Ru interchange by constructing an appropriate supercell (see Methods). Such an interchange results in two types of *defects*. The first kind of defects are hexagons with a Ru instead of a Li atom at the centre and the second kind of defects are Ru dimers broken by the substitution of Li atoms. We found that this breaking up of the Ru dimers leads to significant changes in the magnetic properties of the system. In an ionic model, unpaired magnetic $Ru^{4+}$ ions would have S = 1 and hence a magnetic moment of $2\mu_B$, while our GGA calculations show the magnetic moment of ~$1.2\mu_B$ on this unpaired Ru atom. In addition, there are two other Ru ions next to the Li ion breaking the dimer, which themselves are magnetized by this defect with the induced magnetic moments of ~$0.7\mu_B$. We note that the total change of the magnetic moment with the Li/Ru interchange found in our calculations is ~$0.25\mu_B$/f.u., consistent with our experimental findings. The difference in the spin densities for pure and 4% doped $Li_2RuO_3$, shown in Fig. SI10, clearly demonstrates a formation of the magnetic cloud in a vicinity of defects. Thus the obtained theoretical results support the picture of short-scale magnetic cloud close to the impurity, as described above.

**Summary**

To summarize, our detailed and extensive studies on the disorder effects of $Li_2RuO_3$ paint two apparently contradicting, yet rather revealing pictures of the spin dimerization. The Ru−Ru singlet dimers that form a long-range ordered state (valence bond crystal) for pure $Li_2RuO_3$ below $T_c$ ~ 540 K, are quite robust. And yet, at the same time, such valence bond crystal itself appears to be very fragile and can be suppressed by rather low levels of disorder. Thus the state of $Li_2RuO_3$ with a finite level of disorder seems to correspond not to a valence bond crystal, but rather to a valence bond liquid (or valence bond glass), similar in spirit to the resonating valence bond state proposed originally for frustrated magnets[27]. Our work shows that $Li_2RuO_3$ is a very convenient and interesting material, providing good playground, on which one can test and improve our understanding of the unique transition from a valence bond crystal state to a valence bond liquid state, a quite nontrivial quantum state of matter, and eventually to a magnetically ordered state. Thus, it provides a rare window of opportunity to study the question of the destruction of quantum-entangled states in a real material.

**Acknowledgements**

We thank J. Park, H. Jung, and J. Taylor for their help with some of the experiments. We acknowledge D. C. Ahn for technical assistance during our experiments at the Pohang Accelerator Laboratory, and W. Kockelmann for his help on GEM measurements. We also thank S. Nagler, G. Balakrishnan, A.D. Hillier, Y. B. Kim, J. T. Chalker, Z.V. Pchelkina, M.A. Korotin and E.K. Moon for useful discussions. The work at the IBS CCES was supported by the research programme of Institute for Basic Science (IBS-R009-G1). SS acknowledges the support by Russian Scientific Foundation via program 14-22-00004, and the work of D.Kh. was supported by the Koeln University via German excellence initiative and by the German project FOR 1346. DTA would like to thank CMPC-STFC (grant number CMPC-09108) for financial support.

**Authors Contributions**

JGP conceived and supervised the project: DTA was in charge of neutron scattering experiments and related works at ISIS, UK. JHP and TYT were responsible for synthesis of all the samples and bulk characterization at SNU while SIC, DYC, SHL, JYK, HSS, MRL were involved in some of the data collection and sample characterization: PM was involved in the sample synthesis at ISIS. JHP, TM, and HN made the magnetoresistance measurements. Neutron scattering experiments were proposed, carried out and analyzed, including interpretation, by DTA and VK: ADA performed neutron diffraction experiments and analyzed the data: SS carried out GGA calculations. JGP and DIK worked out the interpretation of the data and wrote the manuscript after discussion with SS and DTA.

**Additional Information**

**Supplementary information** accompanies this paper at http:xxxxx
**Competing financial interests:** The authors declare no competing financial interests.

**Methods**

Sample preparation & physical properties

In order to make comprehensive studies of doping experiments on the dimerized ground state of $Li_2RuO_3$, we prepared 7 samples with different levels of disorder using a solid state reaction method. All our samples were made with the starting materials better than 99.99% purity by mixing stoichiometric amounts of Ru or $RuO_2$ and $Li_2CO_3$, preheated at 600 °C in air for overnight. The powder was then pressed into 10 mm diameter pellets and fired at 900 °C for 15 hr, before being further sintered in air at 1000 °C from 48 to 200 hr with intermediate grinding. The exact final sintering condition of our samples, all labelled with the prefix LRO, is summarized in Table SI2. In addition, one sample labelled DTA was also synthesized at ISIS, UK, by mixing commercially available powders of $RuO_2$ and $Li_2CO_3$ following previously described procedures[14]. The starting raw materials were sintered in air at 1000 °C for 24 h in an alumina crucible, after which the product was reground and pelletized and heated at 900 °C for 48 h.

As well as noting the nominal starting composition of the six (LRO) samples, the exact disorder or mixing ($x$) of Li and Ru content in these samples was determined from an analysis of our high-resolution neutron diffraction data (Supplementary Fig. 4 & Table 4). The values of $x$ obtained suggest that at these low levels of doping the unit cell volume is approximately linear to $x$ as shown in Fig. 3. The sample purity was monitored by collecting powder x-ray diffraction patterns (using a Miniflex II, Rigaku) with Cu $K_\alpha$ radiation. The local structure of our samples was examined by using the XANES (X-ray absorption near edge structure) technique at the 3C1 beam line of the Pohang Accelerator Laboratory and further high-resolution x-ray diffraction experiments were conducted at 8C2 beam line of the Pohang Accelerator Laboratory and using a high-resolution X-ray diffractometer (or Bruker XRD D8 Discover diffractometer) (Supplementary Figs. 2 & 3). We also carried out microprobe chemical analysis by using ICP (inductively coupled plasma) & EPMA (electron probe micro-analyzer) techniques, and confirmed the chemical variations as discussed in the text.

We carried out low and high temperature resistivity measurements using a home-made setup covering the temperature range from 3 to 650 K. We also measured magnetization and heat capacity measurements using commercial set-ups: MPMS-5XL (Quantum Design, USA), PPMS9 (Quantum Design, USA), and VSM (Lakeshore Instrument). Magnetoresistance measurements were made up to 14 tesla using a home-made set-up at Tohoku Univ.

Neutron scattering

In order to investigate the structural transition in detail, we have also undertaken high-resolution neutron diffraction studies on two samples (LRO2 with $x\sim0.07$ & DTA with $x\sim0.13$) with slightly different levels of disorder by employing two neutron powder diffraction beam lines (HRPD and GEM) at the ISIS, UK from 300 to 590 K. We also measured the spin dynamics of three samples (LRO2, DTA & LRO5 with $x\sim0.16$) at the MERLIN and MARI time-of-flight inelastic neutron scattering beam lines of ISIS, UK from 5 to 580 K.

GGA calculations

The *ab-initio* calculations were carried out by pseudopotential method in the Quantum Espresso code[28]. We used the Generalized Gradient Approximation (GGA) with exchange-correlation potential as proposed by Perdew, Burke, and Ernzerhof[29]. The cut-off energies for the wave functions and charge density were chosen to be 40 and 180 Ry, respectively. The crystal structure was taken from Ref. 14 for $T = 300$ K. In order to simulate the Li/Rh interchange we constructed supercell, consisting of 144 atoms in the same hexagonal plane.

**Figure captions**

Figure 1 **Schematic diagram and in-plane view of the structure**: (a) Top and (b) side view of the crystal structure of P2$_1$/m space group. The Ru honeycomb network changes from (c) a structure at 550 K with the *C2/m* space group to (d) a structure with Ru dimer formation at 300 K with the *P2$_1$/m* space group with the Ru orbital wave function as shown in (e).

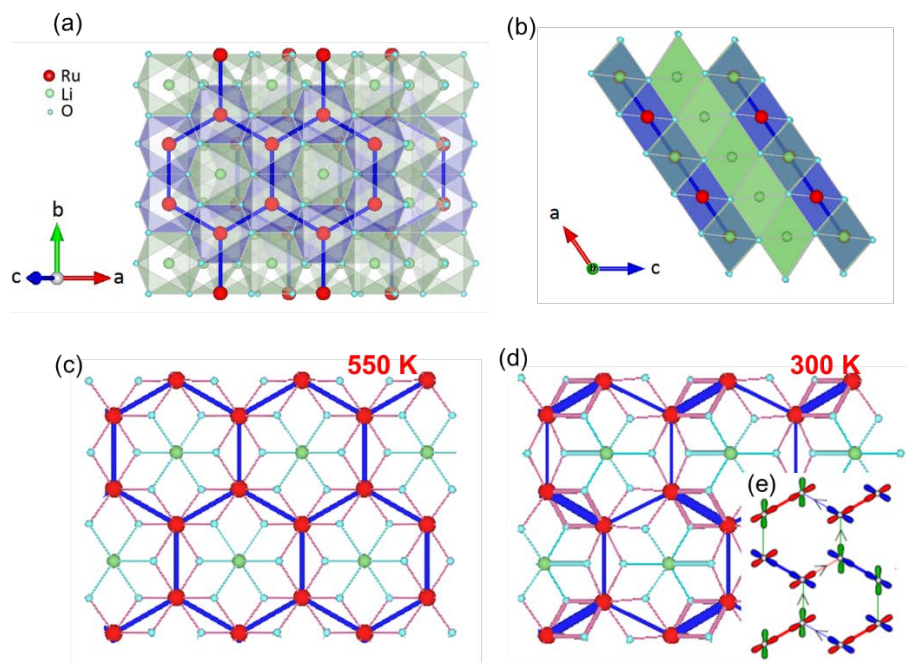

Figure 2 **Bulk properties of our samples**: (a) Resistivity and (b) magnetization, and (c) the low-temperature specific heat for seven $Li_2RuO_3$ samples with different mixing ratio ($x$) values together with $Li_3RuO_4$ and $Li_2TiO_3$. The dashed lines in (a) represent the fitting results using the activation formula. The insets in (a) and (b) show the resistivity and susceptibility data versus temperature for three representative samples, respectively.

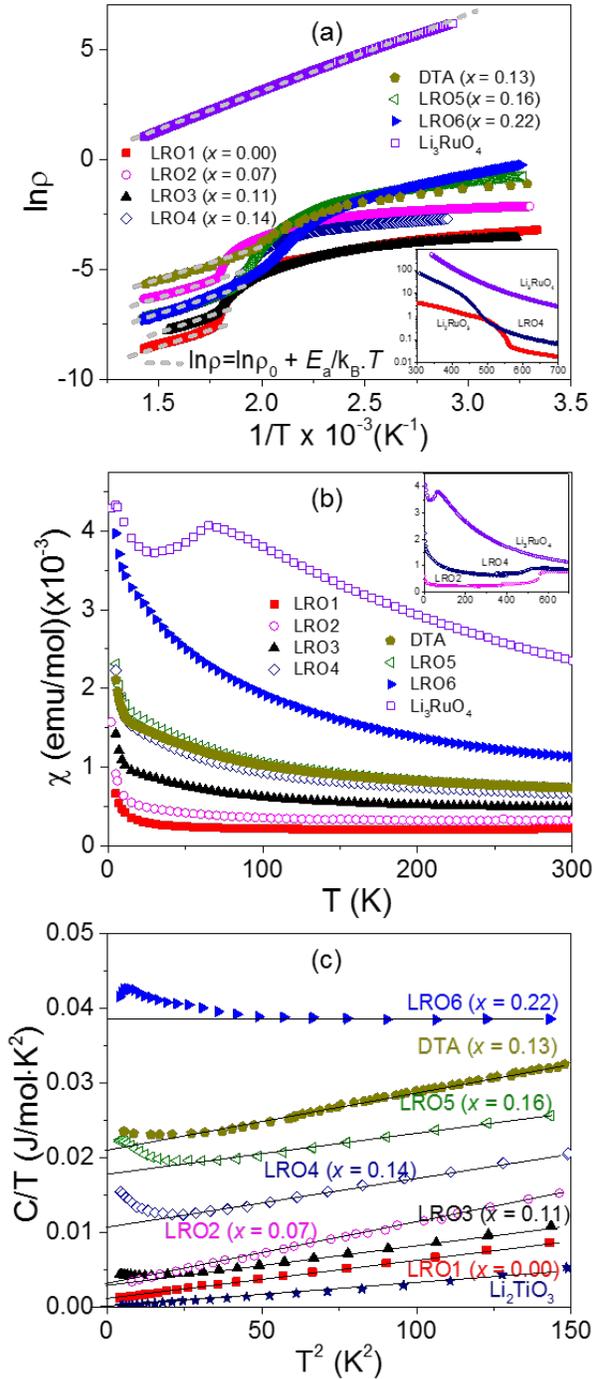

Figure 3 **Doping dependence of key experimental parameters**: (a) the MIT transition temperature, (b) the charge gap estimated from the resistivity data above the transition, (c) the linear temperature dependence to the specific heat and (d) the paramagnetic contribution of the susceptibility.

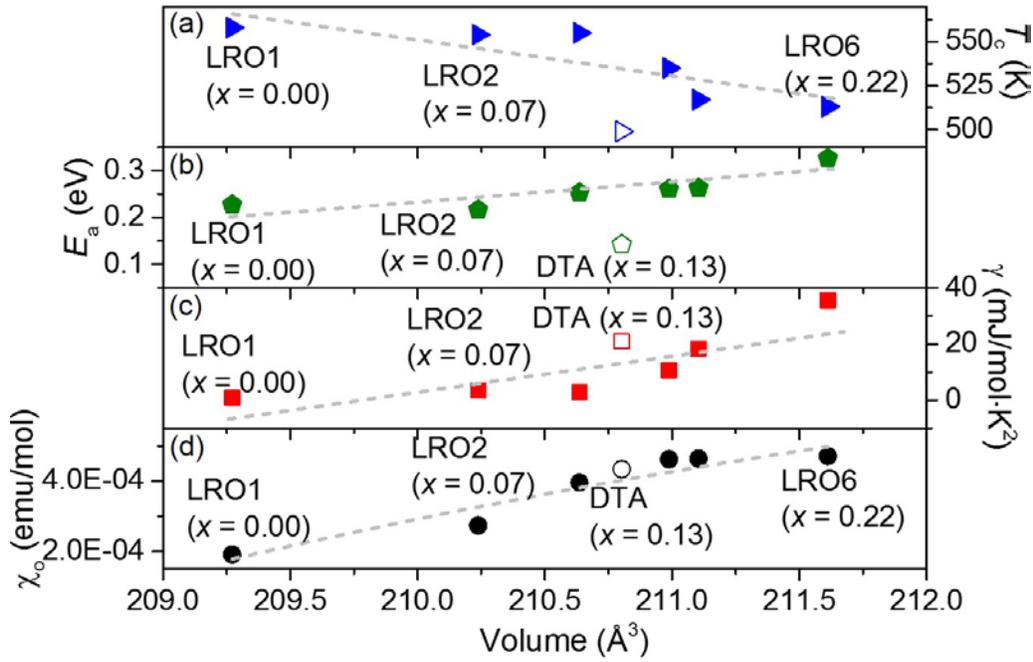

Figure 4 **Spin dynamics measured by inelastic neutron scattering of two Li$_2$RuO$_3$ samples**: (a) DTA ($x = 0.13$) & (b) LRO2 ($x = 0.07$), (c) the difference (DTA-LRO2) plot and (d) the momentum average scattering response as a function of energy for both samples and (e) the difference plot of the momentum average data. We fitted the difference data in (e) using two Lorentzian functions (dash-dot line) with the sum of the two given in the solid line.

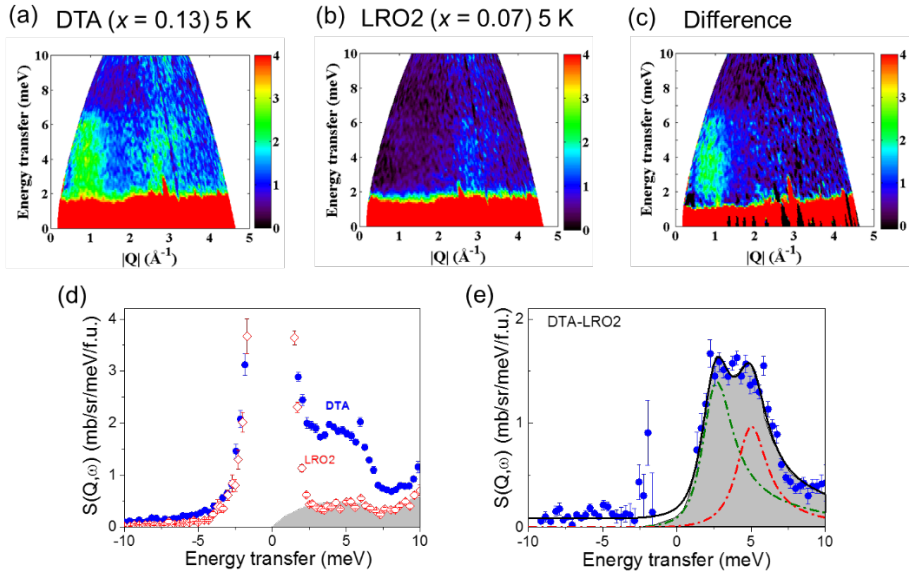

# Robust singlet dimers with fragile ordering in two-dimensional honeycomb lattice of Li$_2$RuO$_3$

Junghwan Park*[1], Teck-Yee Tan*[2], D. T. Adroja[3,4,$], A. Daoud-Aladine[3], Seongil Choi[1,2], Deok-Yong Cho[1,2], Sang-Hyun Lee[1,2], Jiyeon Kim[1], Hasung Sim[2,5], T. Morioka[6], H. Nojiri[6], V. V. Krishnamurthy[7], P. Manuel[3], M. R. Lees[8], S.V. Streltsov[9,10], D.I. Khomskii[11,$], and Je-Geun Park[1,2,5,$]

[1] Center for Strongly Correlated Materials Research, Seoul National University, Seoul 08826, Korea
[2] Center for Correlated Electron Systems, Institute for Basic Science, Seoul 08826, Korea
[3] ISIS Facility, Rutherford Appleton Laboratory, Didcot OX11 0QX, United Kingdom
[4] Highly Correlated Matter Research Group, Physics Department, University of Johannesburg, Auckland Park 2006, South Africa
[5] Department of Physics and Astronomy, Seoul National University, Seoul 08826, Korea
[6] Institute for Materials Research, Tohoku University, Sendai 980-8577, Japan
[7] Department of Physics and Astronomy, George Mason University, Fairfax, VA 22030-4444, USA
[8] Department of Physics, University of Warwick, Coventry CV4 7AL, United Kingdom
[9] Institute of Metal Physics, Ekaterinburg 620041, Russia
[10] Department of Theoretical Physics and Applied Mathematics, Ural Federal University, Ekaterinburgh 620002, Russia
[11] II Physikalisches Institut, University of Koeln, 50937 Koeln, Germany

* Equally contributed
$ Correspondence to J.G.P. [email: jgpark10@snu.ac.kr], D.I.K. [email: khomskii@ph2.uni-koeln.de] & D.T.A. [email: devashibhai.adroja@stfc.ac.uk]

SI Table 1: Summary of the structure analysis of the Li$_2$RuO$_3$ sample with the data given in Fig. SI1.

| Sample | LRO3 | |
| --- | --- | --- |
| Temperature (K) | 300 | 550 |
| Space group | $P2_1/m$ | $C2/m$ |
| $a$ (Å) | 4.9392(3) | 5.0461(2) |
| $b$ (Å) | 8.7692(5) | 8.7537(3) |
| $c$ (Å) | 5.8881(3) | 5.9287(2) |
| $\beta$ (°) | 124.4446(25) | 124.5277(31) |
| $V$ (Å$^3$) | 210.333(21) | 215.752(15) |
| Ru–Ru (Å) | 2.605(4)<br>3.012(4)<br>3.052(2) | 2.934(2)<br>2.907(1)<br>2.907(1) |

SI Table 2: Summary of the detailed final sintering conditions for the six $Li_2RuO_3$ samples with different mixing ratio ($x$) between Li and Ru atoms on the Ru honeycomb lattice.

| Sample Name | Mixing ratio ($x$) | Starting materials | Sintering condition |
|---|---|---|---|
| LRO1 | $x \approx 0$ | $Li_2CO_3$ (10 mol % excess) + Ru | 1000 °C for 98 h (pellet placed in alumina crucible) |
| LRO2 | $x \approx 0.07$ | $Li_2CO_3$ (10 mol % excess) + $RuO_2$ | 1000 °C for 48 h (pellet placed in alumina crucible) |
| LRO3 | $x \approx 0.11$ | $Li_2CO_3$ (10 mol % excess) + $RuO_2$ | 1000 °C for 92 h (pellet placed in alumina crucible) |
| LRO4 | $x \approx 0.14$ | $Li_2CO_3$ (10 mol % excess) + Ru | 1100 °C for 198 h (powder inserted in Pt tube, sealed tightly) |
| LRO5 | $x \approx 0.16$ | $Li_2CO_3$ (10 mol % excess) + $RuO_2$ | 1000 °C for 48 h (pellet placed in alumina crucible) |
| LRO6 | $x \approx 0.22$ | $Li_2CO_3$ (20 mol % excess) + $RuO_2$ | 1000 °C for 48 h (pellet placed in alumina crucible) |

SI Table 3: Summary of the structural data for the six $Li_2RuO_3$ samples with different mixing ratio ($x$). The powder x-ray diffraction patterns are shown in Fig. SI3. A comparison of the FWHM(full width at half maximum) of the (001) Bragg peak for all six samples with that for $Li_2TiO_3$ indicates that all our samples form with a very high level of crystallinity.

| Sample | FWHM (degree) | $a$ (Å) | $b$ (Å) | $c$ (Å) | $\beta$ (°) | $V$ (Å$^3$) |
|---|---|---|---|---|---|---|
| LRO1 | 0.148(6) | 4.9245(3) | 8.7665(6) | 5.8880(4) | 124.382(1) | 209.272(23) |
| LRO2 | 0.164(5) | 4.9217(2) | 8.7788(4) | 5.8938(2) | 124.352(2) | 210.238(15) |
| LRO3 | 0.168(8) | 4.9334(3) | 8.7790(5) | 5.8946(4) | 124.406(3) | 210.636(22) |
| LRO4 | 0.174(6) | 4.9426(2) | 8.7818(3) | 5.8934(3) | 124.435(2) | 210.988(16) |
| LRO5 | 0.173(7) | 4.9440(3) | 8.7826(4) | 5.8957(4) | 124.450(3) | 211.104(21) |
| LRO6 | 0.172(6) | 4.9658(4) | 8.7764(7) | 5.8946(5) | 124.540(5) | 211.612(29) |
| $Li_2TiO_3$ | 0.166(6) | 5.0713(4) | 8.7882(6) | 5.0909(3) | 109.294(8) | 214.143(26) |

SI Table 4: Summary of the structural analysis of two $Li_2RuO_3$ samples with different mixing ratio ($x$) from high-resolution neutron diffraction data taken in the high-temperature phase (see Fig. SI 4).

| Sample | LRO2 | DTA |
|---|---|---|
| Mixing ratio ($x$) | 0.066(2) | 0.129(4) |
| Temperature (K) | 550 | 580 |
| Space group | $C2/m$ | $C2/m$ |
| $a$ (Å) | 5.05099(4) | 5.06358(7) |
| $b$ (Å) | 8.76479(8) | 8.75096(11) |
| $c$ (Å) | 5.94204(7) | 5.93822(9) |
| $\beta$ (°) | 124.530(1) | 124.614(1) |
| $V$ (Å³) | 216.715(4) | 216.556(5) |
| Ru–Ru (Å) | 2.934(4) 2.911(2) | 2.933(6) 2.914(3) |

SI Fig. 1: Structure refinement of the Li$_2$RuO$_3$ sample (LRO3) below and above the transition temperature using high-resolution XRD (X-ray diffraction) data with the summary given in Table SI 1. The green ticks indicate the position of the Bragg peaks and the blue lines at the bottom show the difference curves. The insets show the enlarged pictures of the data at low angles.

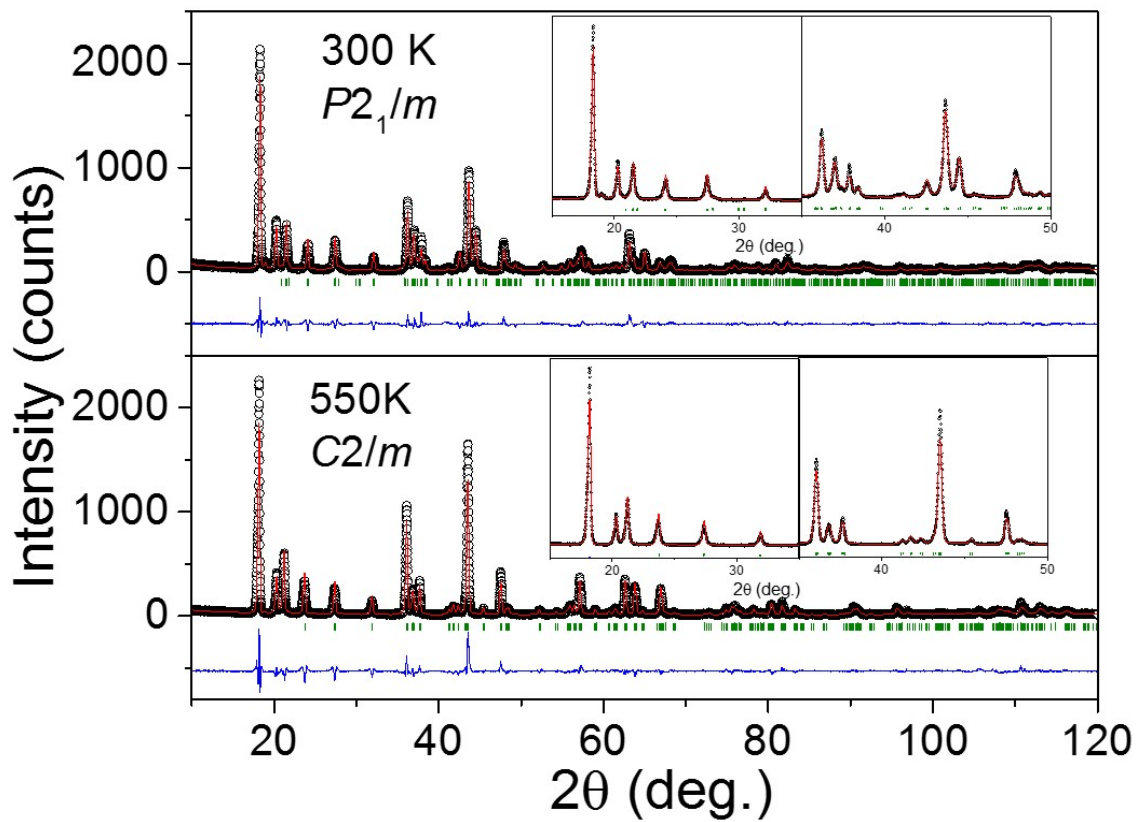

SI Fig 2: High resolution x-ray diffraction data of $Li_2RuO_3$ and $Li_3RuO_4$. The symbols represent the data while the red line is the refinement results. The green ticks indicate the position of the Bragg peaks and the blue lines at the bottom show the difference curves.

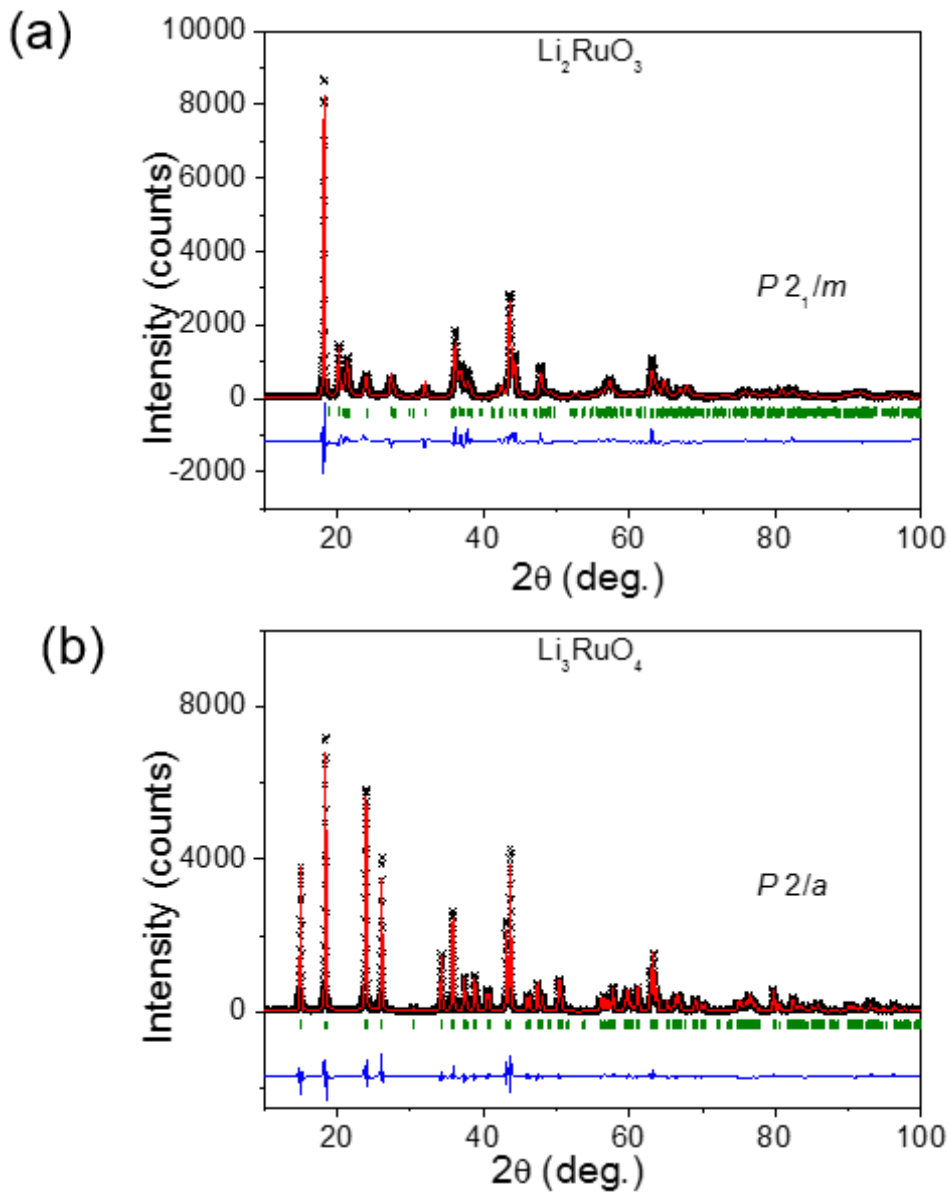

SI Fig 3: X-ray diffraction data of six Li$_2$RuO$_3$ samples with different mixing ratio ($x$). The synthesis methods are summarized in Table SI 2. The shaded area indicates where we expect to see the superlattice peak of the $P2_1/m$ phase.

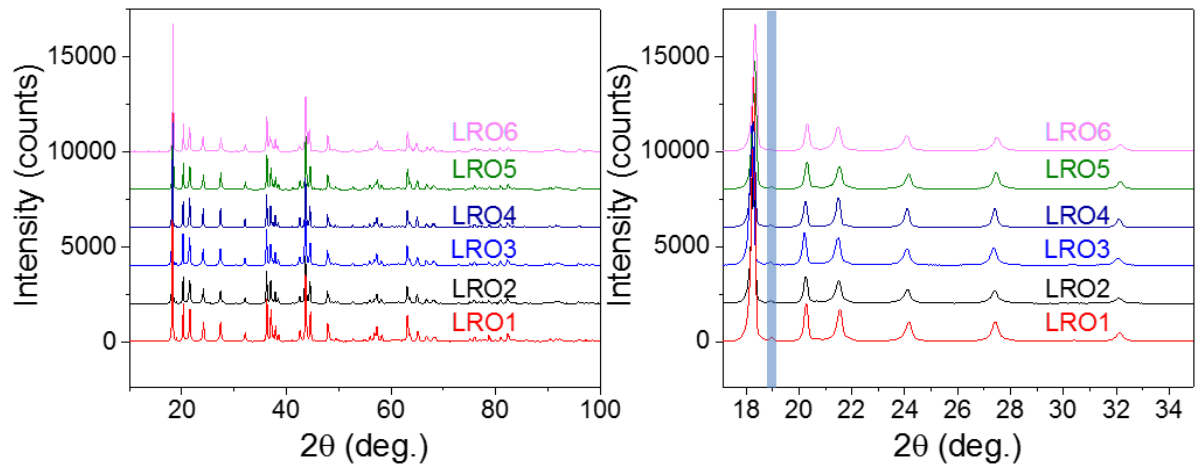

SI Fig 4: Neutron diffraction data taken at high temperature for two $Li_2RuO_3$ samples: LRO2 (with $x = 0.07$) and the DTA (with $x = 0.13$) samples. The green ticks indicate the position of the Bragg peaks and the blue lines at the bottom show the difference curves. A summary of the refinement results is given in Table SI4.

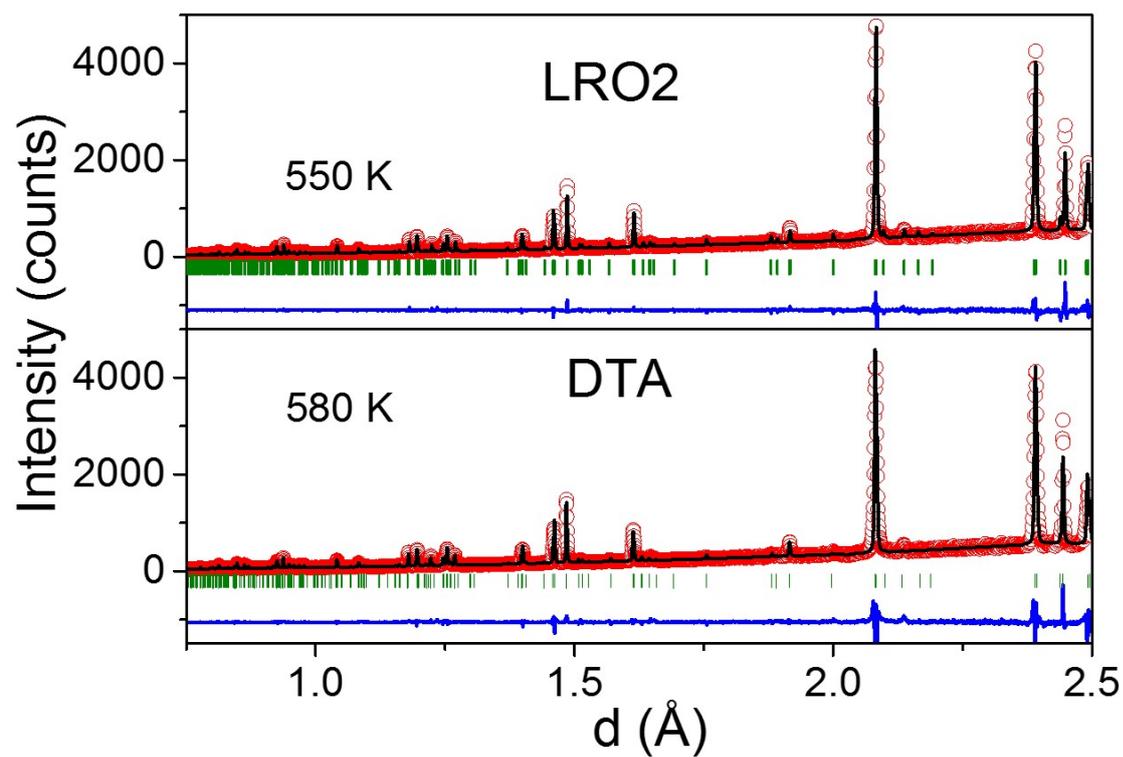

SI Fig 5: (a) Colour plot of scattering intensity as a function of energy transfer & Q and (b) Q-integrated energy cut for the Q range from 0 to 1.5 Å$^{-1}$, measured by an inelastic neutron scattering technique on the Li$_2$RuO$_3$ sample: LRO5 (with $x = 0.16$). (c) The heat capacity and (d) susceptibility data shown in the bottom two panels demonstrate that this new LRO5 sample (2$^{nd}$ batch) synthesized for the inelastic neutron experiments has almost the same bulk properties as the LRO5 sample (1$^{st}$ batch) that was used for the bulk measurements shown in Figs. 2 & 3.

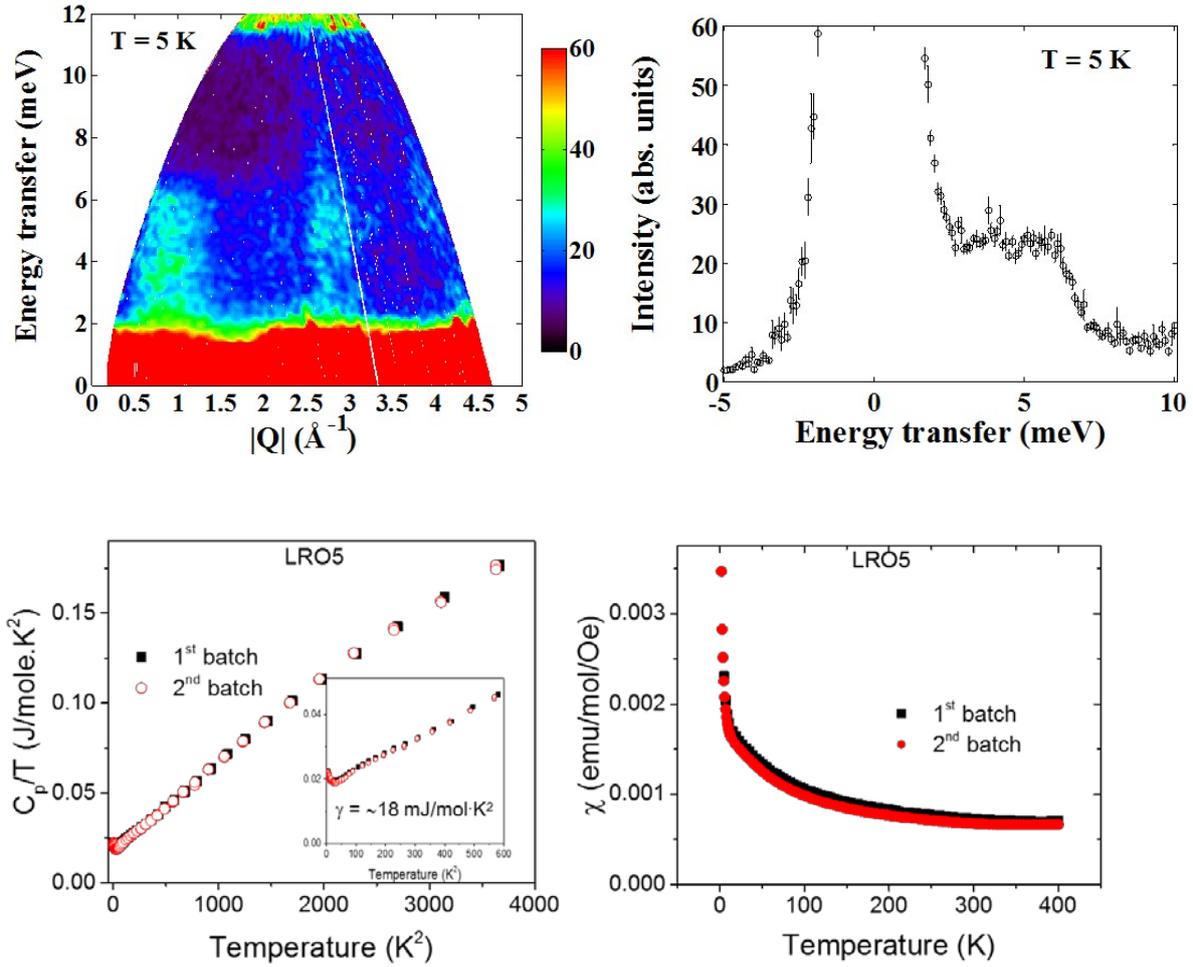

SI Fig 6: Temperature dependence of the low-energy magnetic excitations measured on two Li$_2$RuO$_3$ samples: LRO2 ($x = 0.07$) and DTA ($x = 0.13$) with an incident energy Ei=13 meV at MERLIN.

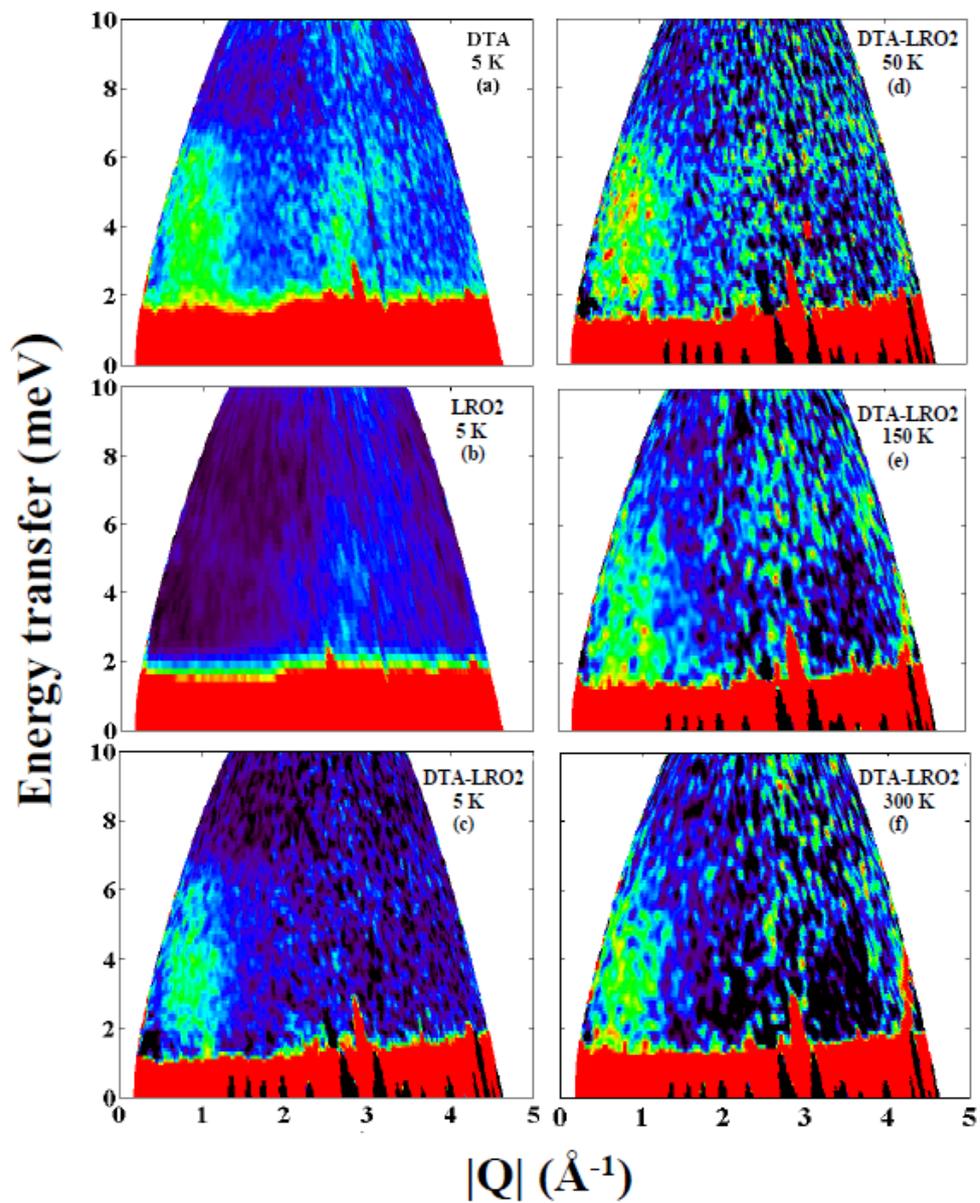

SI Fig 7: Energy versus momentum plot of (a) the low-energy magnetic excitations and (b) their Q-dependence of the $Li_2RuO_3$ DTA ($x$ = 0.13) sample. The solid lines in (b) show the fit to the isolated dimer model using different values of correlation lengths while the dotted line represents the background.

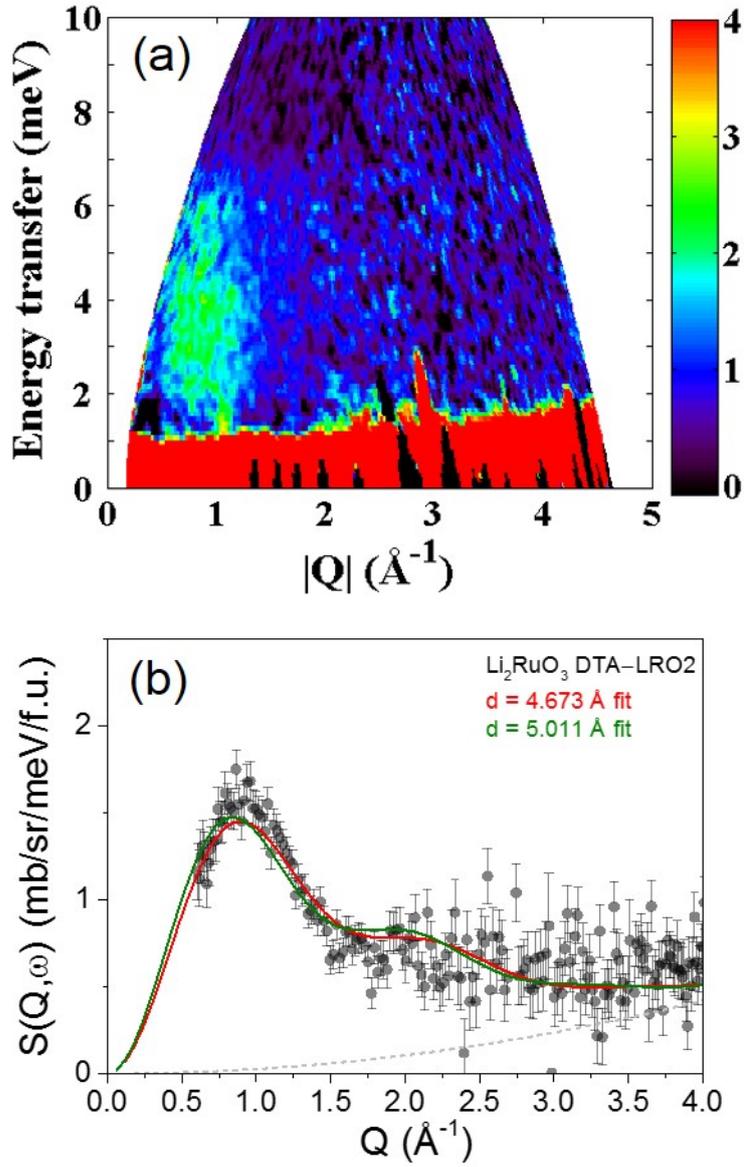

SI Fig 8: Comparison of the bulk susceptibility with the uniform susceptibility calculated using the inelastic neutron scattering data for the two $Li_2RuO_3$ samples (LRO2 with $x = 0.07$ & DTA with $x = 0.13$).

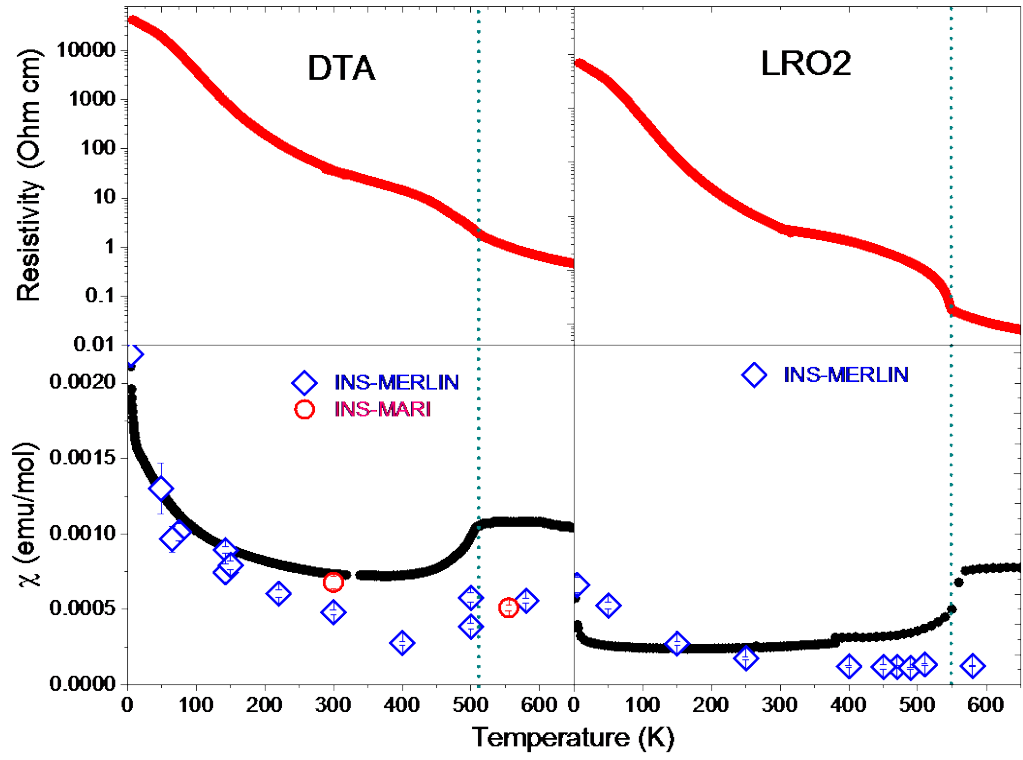

SI Fig 9: A plot of the electronic contribution to the heat capacity (γ) versus the low-temperature value of the susceptibility ($\chi_0$) for the $Li_2RuO_3$ systems with different levels of disorder ($x$) compared with those for other heavy fermion systems. Data for the other heavy fermion systems were taken from Ref. SI 1. This plot shows that the Sommerfeld-Wilson ratio: $R = \frac{4\pi^2 k_B^2 \chi_0}{3(g\mu_B)^2 \gamma}$, is found to be less than one for our $Li_2RuO_3$ materials with different levels of disorder ($x$).

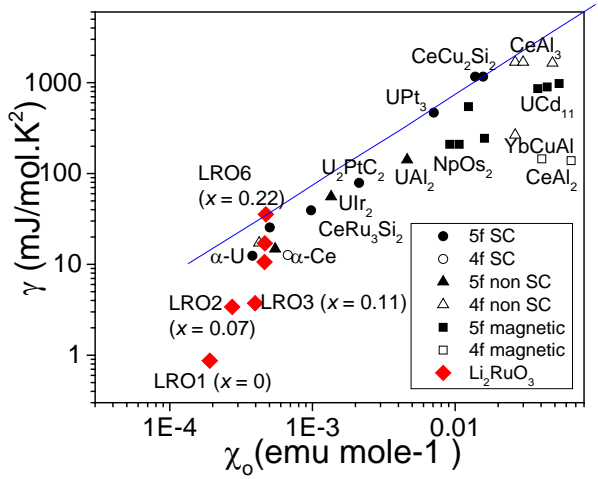

SI Fig 10: A plot of the difference between spin-densities in normal and Li/Rh interchanged Li$_2$RuO$_3$. By colours we show different signs of this difference. Ru ions are grey balls, connected by thick line if they form a dimer. O and Li are not shown for simplicity. One can see one broken dimer (bottom left side) and the hexagon with Ru in its centre (top right side).

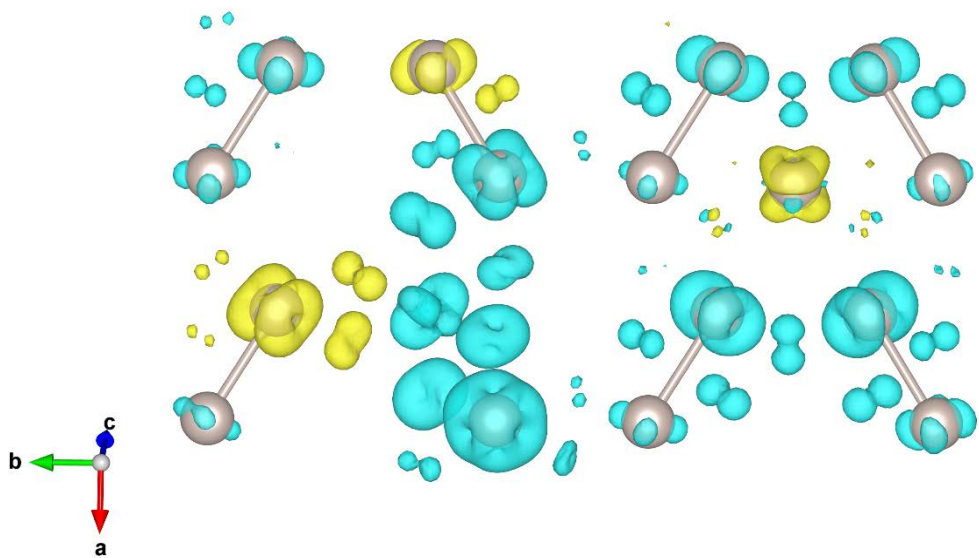